\documentclass[10pt,a4paper,hidelinks,twocolumn]{article}

\usepackage[utf8]{inputenc}

\usepackage{mathpazo}
\usepackage[light]{FiraSans}

\usepackage[sc,sf,bf]{titlesec}

\usepackage[cm]{fullpage}

\usepackage{upgreek}
\usepackage{textgreek}

\usepackage{amsmath}

\usepackage{graphicx}
\usepackage{subfig}
\usepackage[export]{adjustbox}
\usepackage[dvipsnames]{xcolor}
\usepackage{placeins}
\usepackage[labelsep=period]{caption}
\usepackage[switch]{lineno}

\usepackage{siunitx}
\DeclareSIUnit\rpm{rpm}
\DeclareSIUnit\kVp{kVp}
\DeclareSIUnit\pixel{px}

\RequirePackage{titling}
\pretitle{\sffamily \scshape \Huge \centering \bfseries}
\posttitle{\par\medskip}
\predate{\medskip \sffamily \scshape \large \centering}
\postdate{\par\medskip}

\usepackage{abstract}

\usepackage[style=nature,backend=biber,url=false,isbn=false,date=year]{biblatex}
\AtEveryBibitem{\clearfield{issue}}

\usepackage{hyperref}

\usepackage[noblocks,auth-sc,affil-it]{authblk}

\usepackage[shortcuts]{extdash}

\usepackage{tabu}

\begin{document}

\title{Refraction beats attenuation in breast CT}

\author[1,2,*]{Michał Rawlik}
\author[1,2]{Alexandre Pereira}
\author[1,2]{Simon Spindler}
\author[1,3]{Zhentian Wang}
\author[1,2]{Lucia Romano}
\author[2]{Konstantins Jefimovs}
\author[1,2]{Zhitian Shi}
\author[1,2]{Maxim Polikarpov}
\author[1,2]{Jinqiu Xu}
\author[1,2]{Marie-Christine Zdora}
\author[1,2]{Stefano van Gogh}
\author[4]{Martin Stauber}
\author[2]{Eduardo Yukihara}
\author[2]{Jeppe B. Christensen}
\author[5]{Rahel A. Kubik-Huch}
\author[5]{Tilo Niemann}
\author[6]{Cornelia Leo}
\author[7]{Zsuzsanna Varga}
\author[8]{Andreas Boss}
\author[1,2]{Marco Stampanoni}
\affil[1]{Institute for Biomedical Engineering, ETH Zürich and University of Zürich, Switzerland}
\affil[2]{Paul Scherrer Institute, Villigen, Switzerland}
\affil[3]{Department of Engineering Physics, Tsinghua University, Haidian District, 100080 Beijing, China}
\affil[4]{GratXray AG, Villigen, Switzerland}
\affil[5]{Department of Radiology, Kantonsspital Baden, Switzerland}
\affil[6]{Interdisciplinary Breast Center, Kantonsspital Baden, Switzerland}
\affil[7]{Department of Pathology and Molecular Pathology, University Hospital Zürich, Switzerland}
\affil[8]{Institute for Diagnostic and Interventional Radiology, University Hospital Zürich, Switzerland}
\affil[*]{corresponding author, email: mrawlik@ethz.ch}

\date{\small \today}


\twocolumn[
  \begin{@twocolumnfalse}
    \maketitle
    \begin{abstract}
    \sffamily
      \noindent For a century, clinical X-ray imaging has visualised only the attenuation properties of tissue, which fundamentally limits the contrast, particularly in soft tissues like the breast.
      Imaging based on refraction can overcome this limitation, but so far has been constrained to high-dose ex-vivo applications or required highly coherent X-ray sources, like synchrotrons.
      It has been predicted that grating interferometry (GI) could eventually allow computed tomography (CT) to be more dose-efficient.
      However, the benefit of refraction in clinical CT has not been demonstrated so far.
      Here we show that GI-CT is more dose-efficient in imaging of breast tissue than conventional CT.
      Our system, based on a 70\,kVp X-ray tube source and commercially available gratings,
      demonstrated superior quality, in terms of adipose-to-glandular tissue contrast-to-noise ratio (CNR), of refraction-contrast compared to the attenuation images.
      The fusion of the two modes of contrast 
      outperformed conventional CT for spatial resolutions better than 263\,μm and an average dose to the breast of 16\,mGy, which is in the clinical breast CT range.
      Our results show that grating interferometry can significantly reduce the dose, while maintaining the image quality, in diagnostic breast CT.
      Unlike conventional absorption-based CT, the sensitivity of refraction-based imaging is far from being fully exploited, and further progress will lead to significant improvements of clinical X-ray CT.
    \end{abstract}
    \bigskip
  \end{@twocolumnfalse}
]



\noindent In 2020, breast cancer was the most commonly diagnosed cancer overall, with over two million cases. Among women it makes up \SI{24.5}{\percent} of the cancer cases and \SI{15.5}{\percent} of the cancer-related deaths\supercite{GLOBOCAN2020}.
The prevalence of breast cancer has prompted most developed countries to
establish mammography screening programmes, which have been shown to reduce
mortality\supercite{broeders2012impact,Lauby2015}.
However, the effectiveness of mammography is disputed.
A retrospective study found that only \SI{46}{\percent} of screen-detected cancers are true positives, while \SI{22}{\percent} are missed\supercite{Hovda2021}.

The reason is that mammography images are difficult to read.
Not only does the soft tissue in the breast provide limited X-ray contrast, but also
the complicated morphology of the breast is ambiguous when rendered in a two-dimensional projection.
Even among experienced readers, the agreement in identifying masses is far from perfect ($\kappa=0.67$)\supercite{Lee2017}.
This is despite the painful measure of compressing the breast to
spread it out on the image and make it thinner, so that the contrast can be improved by using lower-energy X-rays.

The shortcomings of other breast-imaging modalities have so far hindered their widespread use.
Digital Breast Tomosynthesis partially removes the tissue-overlap-related ambiguity but it has been shown to provide only an incremental improvement over mammography\supercite{Skaane2018}.
Breast ultrasound plays mainly a supportive role and,
while MRI provides excellent contrast, its resolution is lower than mammography, it cannot visualise microcalcifications and the modality is expensive, time-consuming and uncomfortable.

Dedicated breast CT has recently been introduced in clinical
practice with promising results, summarised in a recent review\supercite{Zhu2022}.
With the volumetric data, the tissue overlap is completely alleviated, and the breast does not need to be compressed.
In the review, its authors identify the major shortcoming of the method to be the near identical
attenuation contrast between breast tumours without microcalcifications and glandular parenchyma.

\begin{figure*} 
  \centering
  \includegraphics{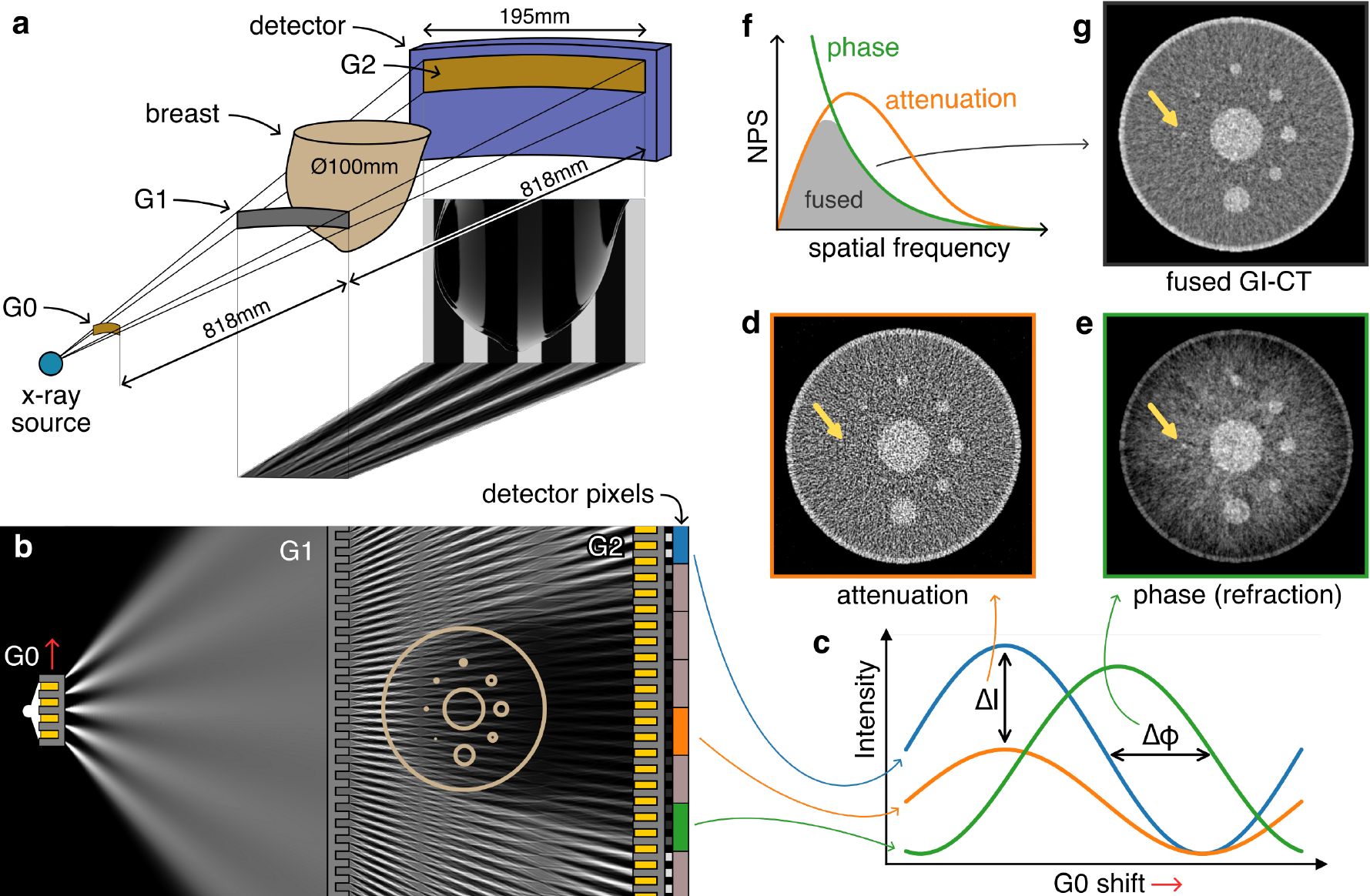}
  \caption{
     \textbf{The principle of GI-CT.}
     \textbf{a} An illustration of the imaging setup.
     The formation of the Talbot interference pattern behind G1 and its distortion by the breast are shown schematically.
     \textbf{b} The intensity of a monochromatic wavefront in a GI-CT system in a horizontal plane.
     The G0 absorption grating acts as an array of vertical slits, each providing enough spatial coherence in the transverse direction so that an interference pattern is created behind the periodically-phase-shifting, by\ \textpi, G1 grating. At the position of G2 the interference produces a pattern of parallel bright-and-dark lines. 
     The period of G0 is chosen such, that the line patterns from its slits add constructively (but incoherently).
     The period of the line pattern is smaller than the pixel size of the detector, so an absorption grating G2, with the same period as the line pattern, is used to analyse it.
     The refraction of X-rays in the sample distorts the line pattern, which results in an intensity change behind the G2 grating.
     The gratings are shown schematically, and their lamellae are vertical, i.e.\ perpendicular to the plane of the image.
     For illustration purposes the image is stretched and shows only an approximately 600\,μm-wide part of the system.
     \textbf{c} Change in the intensity incident on pixels as G0 is shifted: without interaction with the object (blue), with pure attenuation interaction (orange) and with refraction (green).
     \textbf{d} A slice of a phantom reconstructed from the attenuation signal in a simulated GI-CT measurement. The yellow arrow points to a small feature, not discernible in the attenuation image.
     \textbf{e} The phase reconstruction from the same simulated measurement. The small feature can be recognised, but low-spatial-frequency noise is prominent.
     \textbf{f} GI-CT fusion combines the low spatial frequencies of the attenuation reconstruction, where it has lower Noise Power Spectrum (NPS), with the high ones of the phase reconstruction -- the regime where the phase has lower NPS.
     \textbf{g} The fused GI-CT image. The feature pointed out by the arrow is visible and the low-spatial-frequency noise is suppressed.
  }\label{fig:principle}
\end{figure*}

Like visible light, X-rays are not only attenuated but also refracted when traversing matter.
In the last years, several methods to detect the refraction of X-rays,
the phase contrast (PC), have been developed.
In the context of breast imaging, three are noteworthy.
One is \emph{propagation-based PC}, which does not need additional optical elements, but requires the high spatial coherence of a synchrotron or a specialised laboratory source.
The other two, \emph{edge illumination}\supercite{Olivo2021} and \emph{grating interferometry} (GI)\supercite{Pfeiffer2006}, use optical elements, and work with standard X-ray tube sources with large spot sizes.
We discuss GI further, and for a broad overview of PC imaging we direct the reader to a recent review\supercite{Momose2020}.

The principle of GI is illustrated in Fig.\,\ref{fig:principle}a~and~b.
Let us assume for the moment that the G0 element is a narrow slit that provides spatial coherence for the X-ray beam.
G1 is a grating that introduces a periodic $\pi$-shift in the beam, which results downstream in an interference pattern of parallel bright and dark lines.
Refraction on large-scale structures in the sample, the phase contrast, causes the pattern to shift laterally.
Diffusion by refraction on small-scale structures beyond the resolving power of the detector, the \emph{dark-field} (DF) signal, blurs the line pattern.
For X-rays, the refraction angles are in the microradian range requiring the period of the pattern (and G1) to be in the order of a few micrometres.
A pattern this small cannot be resolved directly with standard large-scale detectors, so a periodically opaque analyser G2 with the period matching the one of the pattern is used.
The final observation is that G0 can, in fact, also be a periodically-opaque grating: an array of slits with the spacing chosen so that the line patterns created by each add constructively (but incoherently) in the plane of the analyser\supercite{Pfeiffer2006}.

The gratings are decisive when it comes to the sensitivity of GI to the refraction of X-rays.
The minimal detectable refraction angle is proportional to the pitch of G2, favouring pitch sizes in the few-micrometre range, as well as to the contrast in the interference pattern analysed by G2, called the visibility\supercite{Modregger2011Sensitivity}.
To periodically block X-rays with the energy in the clinical regime, heavy-element, like gold, lines of \SI{100}{\micro\metre}--\SI{200}{\micro\metre} thickness are necessary.
Fabrication of these high-aspect-ratio (line-thickness--to--half-pitch--ratio) microstructures with sufficient quality is challenging.
Deep X-ray lithography (LIGA) can manufacture gratings with thick gold lines (\SI{200}{\micro\metre} but the pitch size is limited to several micrometres\supercite{NodaLIGA, MohrLIGA}.
Silicon-based manufacturing is now pushing the aspect ratio for the pitch size in a micrometre\supercite{Shi2020Micromachines} and sub-micrometre\supercite{Romano2020HighAspectRatio,Romano2020Nanoscale} regimes, for both etching of silicon template and gold filling\supercite{Josell2020Aubottomup}.

The additional refraction information that GI provides is attractive for breast imaging.
PC promises higher contrast\supercite{hellerhoff2019assessment} for better differentiation of tissues and DF was shown to distinguish benign
from malignant calcifications\supercite{scherer2016improved,forte2020can}.
Two-dimensional GI mammography is already close to first clinical
trials\supercite{arboleda2020towards}.

It is natural to pursue the extension of PC to 3D, given the advantages it has shown in two-dimensional imaging.
It has been demonstrated that the benefit of PC-CT depends on the spatial resolution or, for a fixed contrast-to-noise ratio, equivalently, the dose.
It was predicted that with the increase of sensitivity through the progress in the technology of grating fabrication high-resolution breast imaging would be the first clinical area where PC-CT will be advantageous\supercite{raupach2011analytical,Raupach2012}.

So far PC-CT of the breast tissue has been investigated in the high-dose, high-resolution
context of virtual histopathology, demonstrating a clear benefit over attenuation\supercite{vila2021,hellerhoff2019assessment,Massimi2022}.
Propagation-based PC breast imaging is pursued at the Elettra synchrotron in Trieste\supercite{Longo2019} and the Australian Synchrotron\supercite{Arhatari2021}, but the method fundamentally relies on the high coherence of synchrotron X-ray beams and cannot be translated to a wide-spread clinical use.
Small systems for propagation-based PC have been built, but the low power of the required microfocal sources results in very long scan times\supercite{Mettivier2021}.
Edge-illumination was demonstrated for ex-vivo studies of larger specimens\supercite{massimi2021detection}.
A recent review can be found in\supercite{heck2020recent}.
The predicted tipping point of the advantage of clinically-applicable phase contrast has not yet been reported.

We have constructed a GI-CT system
with the aim to demonstrate additional value of GI in clinical breast CT.
We have imaged a formalin-fixed human-breast specimen and compared the dose efficiency of the two GI-CT contrasts, attenuation and phase, for a range of resolutions, as well as the combination of the GI-CT contrast versus conventional, attenuation-based CT.

\section*{Results}

\begin{figure*}
  \centering
  \includegraphics[width=\linewidth]{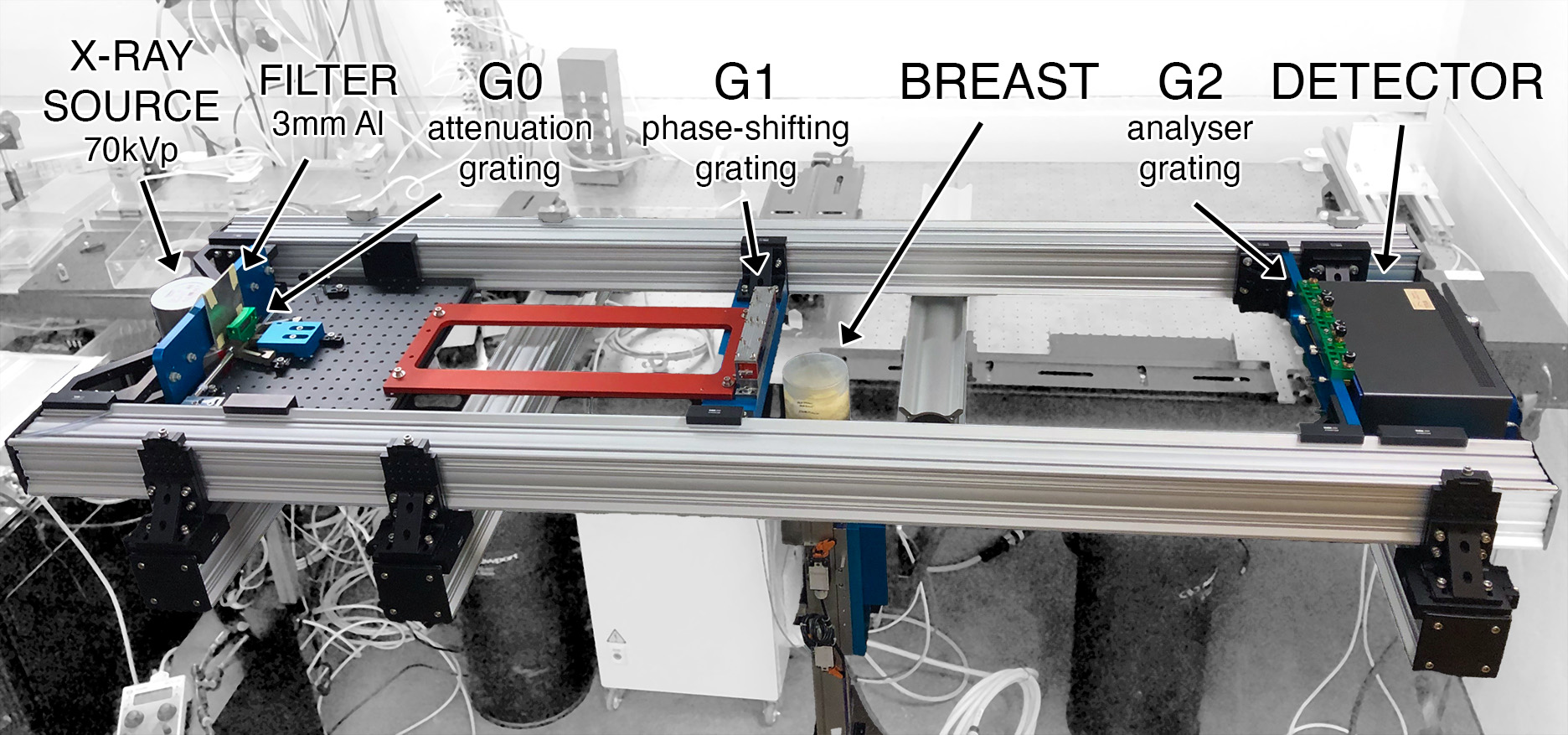}
  \caption{
      \textbf{The GI-CT setup.}
      From the left: the tungsten-anode X-ray source operated at 70\,kVp, 3\,mm Aluminium filter, the G0 attenuation grating, the phase-shifting G1 grating, the breast in a sample holder mounted on a rotation and vertical stages, the array of three G2 analyser gratings, and the detector.
  }\label{fig:setup}
\end{figure*}

The \SI{1.8}{\metre}-long GI-CT system consisted of a tungsten-anode X-ray source operated at \SI{70}{\kVp}, a photon-counting detector with an active area of $195 \times \SI{19.2}{\milli\metre\squared}$ and a
Talbot--Lau interferometer designed for the photon energy of \SI{46}{\keV} in a symmetric geometry, with all gratings being \SI{4.2}{\micro\metre}-pitch.
The system is depicted in Fig.\,\ref{fig:setup} and described in detail in the Methods section.

\begin{figure*}
  \centering
  \includegraphics{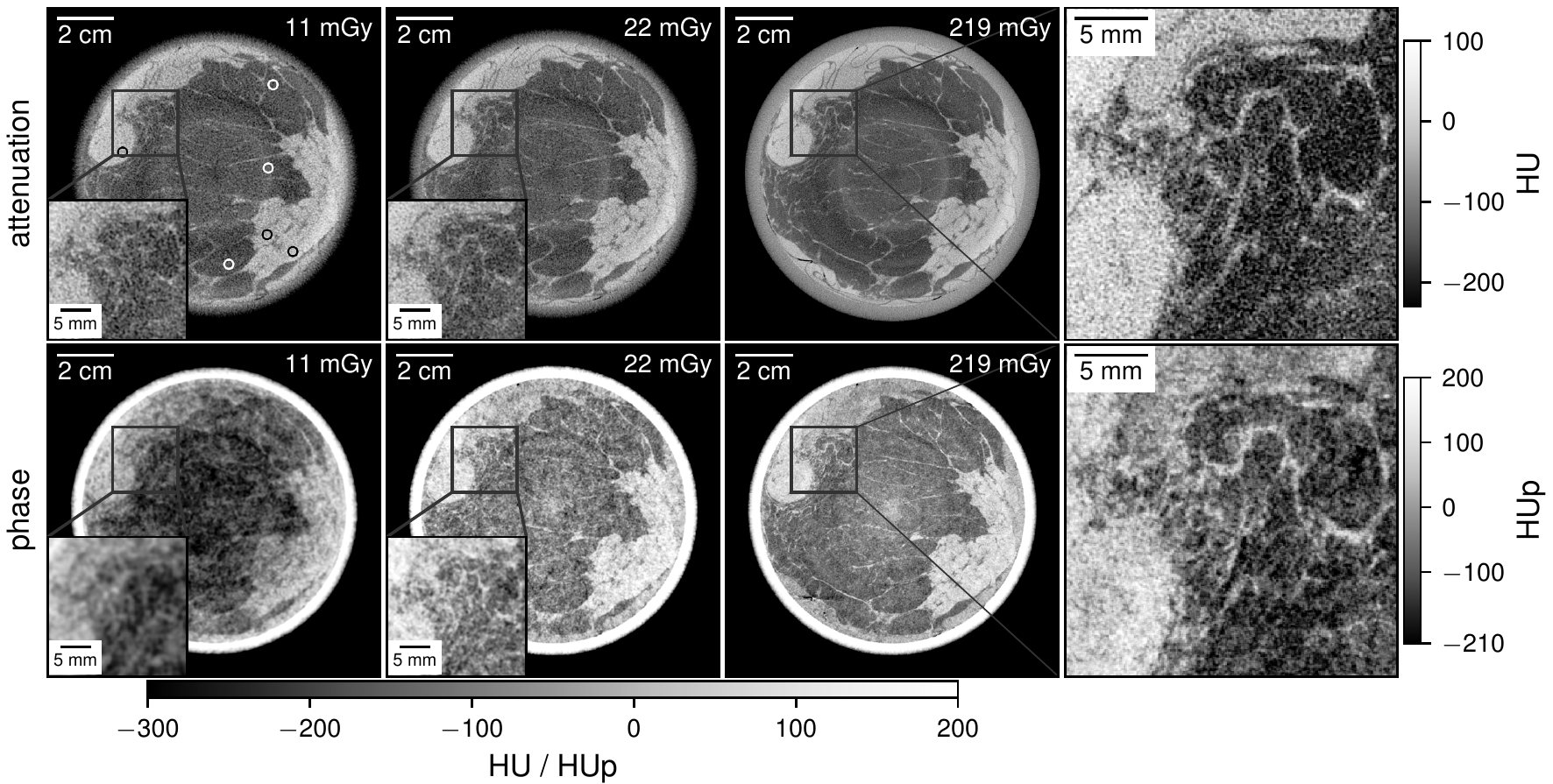}
  \caption{
    \textbf{The axial slices of reconstructed volumes of the human-breast specimen.}\quad
    The attenuation contrast is shown in the top row, the phase contrast in the bottom row.
    Three values of the average dose absorbed in the breast are depicted in the columns: 11\,mGy, 22\,mGy and 219\,mGy.
    Visually, the image quality increases with the dose for both contrasts, but for the phase
    one it increases faster.
    At the dose of 219\,mGy, the image quality of phase contrast is superior, which is visible
    particularly well in the enlarged region in the fourth column.
    In the top-left image, the regions of interest used to calculate the CNR are marked.
  }\label{fig:CT_images}
\end{figure*}

We imaged a formalin-fixed human breast from a body donation with the average dose delivered to the specimen in the range of \SIrange[range-phrase=--]{5.5}{219}{\milli\gray}.
The axial slices of the reconstructed volumes for both attenuation and phase contrast at mean absorbed dose to the breast of \SI{11}{\milli\gray}, \SI{22}{\milli\gray} and \SI{219}{\milli\gray} are shown in~Fig.\,\ref{fig:CT_images}.
The visual quality of both contrasts increases with the delivered dose, which is particularly clear in the insets showing an enlarged portion of the image.
Moreover, even though the phase-contrast image for the lowest dose appears visually inferior to the attenuation-contrast one, for the higher dose it appears superior.

\begin{figure}
  \centering
  \includegraphics{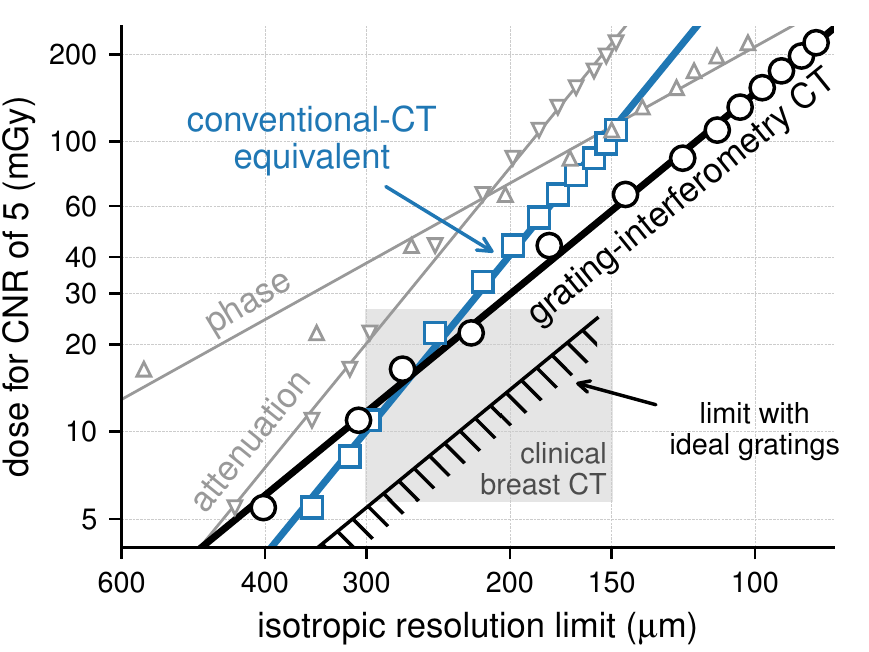}
  \caption{
    \textbf{The average absorbed dose requirement of the conventional CT and GI-CT imaging of the breast as a function of the resolution.}
    Each point represents an image (like in Fig.\,\ref{fig:CT_images}) for which the minimal FWHM size of the isotropic Gaussian kernel to reach the contrast-to-noise ratio (CNR) of 5 was found.
    The kernel size constitutes a lower limit on the resolution.
    The best-fit lines of the attenuation contrast (slope 3.44) and phase contrast (slope 1.56) intersect at the kernel size of 214\,μm and dose of 65\,mGy.
    The ones of the conventional-CT equivalent (the attenuation signal with half the dose, as G2 would not be necessary;
    slope 3.44) and fused GI-CT (slope 2.30) intersect at 263\,μm and 16\,mGy.
    The intersection can be interpreted as the point beyond which GI-CT delivers superior image quality per unit dose.
    Notably, it is within the range of clinical breast CT. 
    The numerically-derived limit for fused GI-CT with ideal gratings, and otherwise the same geometry, is also shown.
    The points for phase below 16\,mGy are omitted for clarity.
  }\label{fig:Relative_CNR}
\end{figure}

We analysed the images quantitatively following an approach inspired by Raupach et al.\supercite{Raupach2012}.
We assumed that, in order to resolve the morphology of the breast, a contrast-to-noise ratio (CNR) of 5 between the adipose and glandular tissue is necessary (the Rose criterion\supercite{rose2013vision}).
With decreasing dose, and consequently increasing noise, a smoother filtering is necessary to reach this CNR value.
The additional point spread function (PSF) of the filtering introduces limits to the imaging resolution. 
For each image, we determined the full-width-half-maximum (FWHM) size of an isotropic Gaussian-blur kernel necessary to achieve the CNR of 5.
The results depicted in Fig.\,\ref{fig:Relative_CNR} can be interpreted as the dose necessary to resolve the morphology as a function of the resolution.
We observed that the dose requirement rises with the power of $3.44$ of the inverse kernel size for attenuation and $1.56$ for phase.
The two curves, having different slopes, intersect at a kernel size of \SI{214}{\micro\metre}, which can be interpreted as the resolution above which, on our system, phase-contrast imaging is more dose-efficient than attenuation.

Due to the differential nature of phase contrast in GI, the reconstructed volumes exhibit different in-plane noise properties than attenuation contrast\supercite{raupach2011analytical}.
The noise power spectrum of PC-CT, in comparison with attenuation CT, is smaller for high spatial frequencies, but larger for low frequencies (Fig.\,\ref{fig:principle}f).
This observation, in addition to the fact that the two contrasts come from a single acquisition, and are therefore naturally registered, motivates fusing the images\supercite{Wang2013ImageFusion}.
Because attenuation and phase carry fundamentally different information there is no generic way to fuse them.
However, in the context of a particular imaging task, which we define to be differentiating glandular and adipose tissue with the maximal possible CNR, we used a simple scheme, illustrated in Fig.\,\ref{fig:principle}f~and~g.
First, we normalised the reconstructed phase-contrast volume so that the grey levels of the adipose and glandular tissues were equal to the ones in the attenuation.
Then, we added the high-pass filtered phase volume to the low-pass filtered attenuation one, both using the same in-plane Gaussian kernel.
We found the optimal size of the kernel, the one maximising the CNR of the fused image, to be close to $\sigma = \SI{1.5}{\pixel}$ for all measurements.
The quantitative CNR analysis of the fused images, depicted in Fig.\,\ref{fig:Relative_CNR}, showed that in the investigated range of kernel sizes they approximately follow a power law with the exponent of $2.30$ lying between the one of the images derived from attenuation and phase contrast.
Moreover, the dose requirement is lower than the ones of each of the single-contrast images everywhere.

\begin{figure}[tb]
  \centering
  \includegraphics[width=\linewidth]{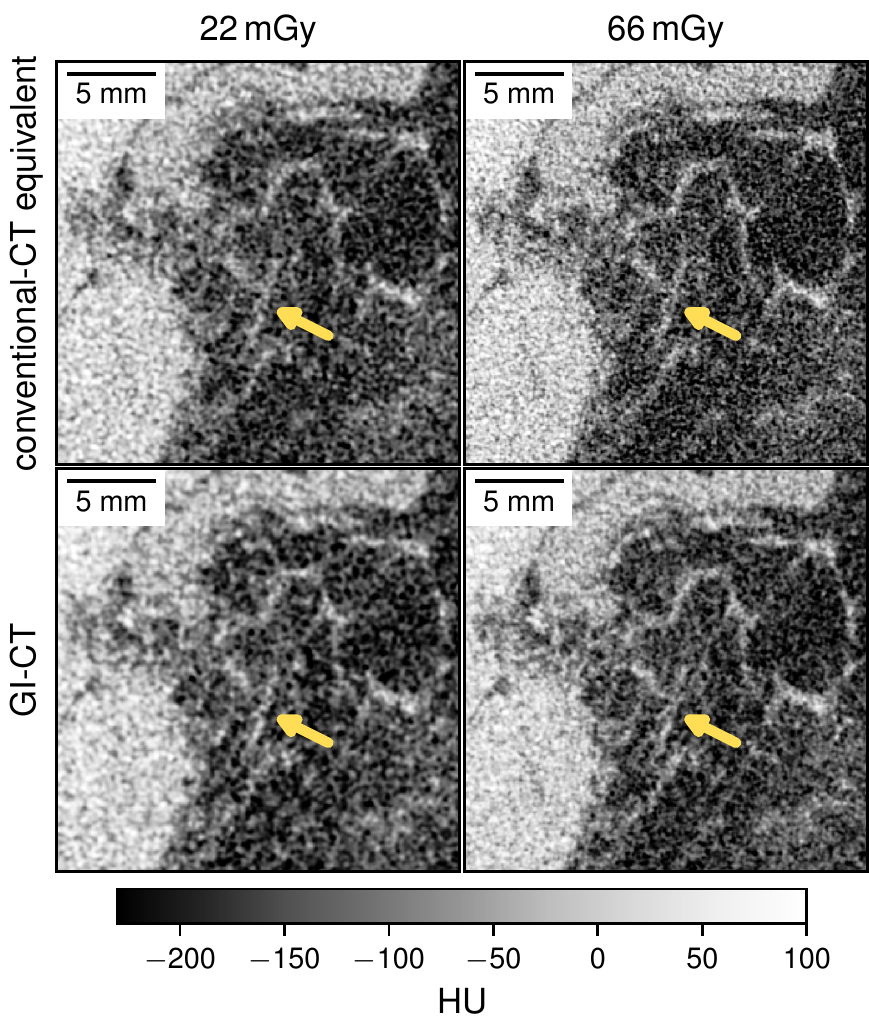}
  \caption{
    \textbf{A comparison of conventional-CT equivalent and fused GI-CT images}\quad
    At the dose of 22\,mGy the additional information coming from refraction allows GI-CT to just about overcome the reduction in statistics coming from the G2 absorbing half of the photon flux and the image quality is comparable to that of a conventional-CT equivalent.
    At 66\,mGy the image quality of GI-CT is superior, which is visible particularly well in small features, like the one indicated by the arrow.
  }\label{fig:CT_images_same_dose}
\end{figure}

Without the G2 grating, which attenuates approximately half of the photon flux downstream of the specimen, it would be possible to acquire an attenuation image with the same photon-counting statistics at half the dose.
We therefore consider the attenuation contrast with half the dose to be approximately equivalent to a conventional CT image.
The comparison of the CNR-dose-efficiency of the fused GI-CT and the conventional-CT equivalent, depicted in Fig.\,\ref{fig:Relative_CNR},  shows that the former outperforms the latter for isotropic kernel sizes sharper than \SI{263}{\micro\metre} and doses larger than \SI{16}{\milli\gray} (for CNR\,=\,5).
In Fig.\,\ref{fig:CT_images_same_dose}, we show a comparison of an enlarged fragment of fused-GI-CT and conventional-CT-equivalent images.
While at the dose of \SI{22}{\milli\gray} the benefit of GI-CT is not visually impressing, it increases with the dose, and it becomes evident at \SI{66}{\milli\gray}, in particular for small features.
We would like to point out that the low-frequency contrast information in the fused image comes from the attenuation contrast and thus for both images the quantitative unit is the one of attenuation, that is Hounsfield's (HU).

\section*{Discussion}
Soon after the first demonstration of grating interferometry with a regular, tube-based X-ray source\supercite{Pfeiffer2006} it has been predicted that the method can greatly improve clinical CT imaging, in particular of the breast\supercite{Raupach2012}.
Despite a decade of intensive research, the practical demonstration has been, so far, elusive.
With the aim to settle the long discussion we have constructed a GI-CT system with a tube-based X-ray source operated at a typical breast-CT energy of \SI{70}{\kVp}, a \SI{10}{\centi\metre}-wide field of view, and a Talbot--Lau interferometer based on commercially-available gratings.
We evaluated its performance by imaging a formalin-fixed human breast specimen.

We could show that the additional information provided by refraction led to two breakthroughs.
Firstly, the phase-contrast images are superior, in terms of adipose-to-glandular tissue CNR per unit dose, than attenuation for kernels sharper than \SI{214}{\micro\metre}.
Previous demonstrations of refraction-based CT imaging of the breast with a large field of view either did not show the benefit over attenuation\supercite{massimi2021detection} or relied on a synchrotron \supercite{Longo2019}, which greatly limits the applicability.

Secondly, even though GI-CT utilises only half of the photon flux compared to conventional CT due to the absorbing analyser grating, the combination of the attenuation- and phase-contrast signals provides sufficient information to compensate the loss already at a kernel size of \SI{263}{\micro\metre} and dose of \SI{16}{\milli\gray} (for CNR of 5), both of which lie in their respective ranges used in the clinics: \SIrange[range-phrase=--]{150}{300}{\micro\metre} and \SIrange[range-phrase=--]{5.8}{26.1}{\milli\gray}\supercite{Berger2019,Zhu2022}.
For sharper kernels GI-CT outperforms conventional attenuation-based CT with increasing benefit, requiring only \SI{53}{\percent} of the dose at \SI{150}{\micro\metre}.

Further improvements are possible.
The limit of the sensitivity of GI-CT to refraction is driven by the microfabrication of the gratings.
Even the currently commercially-available gratings performed already well enough for GI-CT to outperform conventional CT.
The performance limit under an assumption of defect-free gratings, depicted in Fig.\,\ref{fig:Relative_CNR}, suggests that GI-CT could require a factor of 2 to 3 less dose than conventional CT in the range of clinical breast CT.
Improvements in the grating-fabrication technology will take GI-CT closer to that limit and, with smaller grating pitches, possibly beyond it.

Our aim was to compare the phase and attenuation contrasts on an even ground.
Here, we have therefore deliberately kept the analysis to a minimum, refraining from the use of iterative reconstruction algorithms, regularisation and post-processing.
The use of those methods is likely to improve the performance as we quantified it in Fig.\,\ref{fig:Relative_CNR}, and a more specific analysis will be available~\supercite{VanGogh2023}.
In particular, we advice caution in comparing the values with other results obtained with elaborate analysis.
The particularity of these advanced analysis methods and their differences for the phase and attenuation contrasts would, in our opinion, only weaken our otherwise general conclusion: that GI-CT provides fundamentally more information to start with.

We have demonstrated that GI-CT is a new relevant clinical imaging modality, which can be more dose-efficient than conventional CT.
X-ray grating interferometry, unlike other imaging techniques exploiting refraction, is compatible with conventional medical CT scanners\supercite{Viermetz2022} and, therefore, suitable for widespread use in hospitals.
The technique is immediately applicable to dedicated breast CT systems, for which we have shown that it already offers an improvement.
In the future, GI could allow dose reduction in all aspects of clinical CT.


\small
\section*{Methods}

\paragraph*{Measurement setup}
The measurement system consisted of a Comet MXR-225HP/11 tungsten-anode X-ray source operated at \SI{70}{\kVp} and \SI{10}{\milli\ampere} for the \SIrange[range-phrase=--]{22}{222}{\milli\gray} measurements and \SI{2.5}{\milli\ampere} for the \SIrange[range-phrase=--]{5.5}{16.5}{\milli\gray} ones.
The size of the focal spot was measured by the manufacturer to be \SI{250}{\micro\metre} (at \SI{30}{\percent} drop).
The X-ray beam was filtered with a \SI{3}{\milli\metre}-thick aluminium plate.
The images were recorded with a photon-counting detector with \SI{750}{\micro\metre}-thick CdTe sensor and \SI{75}{\micro\metre} pixel size, which was manufactured by Dectris AG, Switzerland. The sensor size was $3072 \times 256$ pixels, but only an area of $2600 \times 256$ pixels could be used.
The Talbot-Lau interferometer was configured in a 5th-Talbot-order symmetric geometry with a G1 designed to introduce a phase shift of $\pi$ at \SI{46}{\kV}. The G0--G1 and G1--G2 distances were both \SI{818.1}{\milli\metre}.
The source--G0 distance was \SI{100}{\milli\metre}.
G0 and G1 were a single piece each, and for G2 three gratings were tiled together.
All gratings were bent around the vertical axis going through the X-ray source's focal spot.
For phase-stepping, G0 was moved with a Physik--Instrumente P-841.6B piezo actuator.
The detector was \SI{1756}{\milli\metre} and the rotation centre \SI{1003}{\milli\metre} away from the source.

\paragraph*{Gratings}
The \SI{4.2}{\micro\metre}-pitch attenuation gratings G0 and G2 had gold lamellae electroplated onto a graphite substrate, and were manufactured with the LIGA process by Microworks~GmbH, Germany.
The gratings had a duty cycle of $0.5$, and gold thickness was in the range of \SIrange[range-phrase=--]{150}{180}{\micro\metre}.
The polymer template was not stripped. 

The $\pi$-shifting \SI{4.2}{\micro\metre}-pitch phase grating G1 was manufactured on a double side polished 8-inch silicon wafer by deep reactive ion etching in a SPTS Rapier system.
A pattern in MEGAPOSIT SPR220-3.0 positive tone photoresist was realised by direct laser writing (Heidelberg DWL66+) (see\supercite{Spindler2022theChoice} for further details).
The process was optimised to ensure uniform etching depth and vertical trench sidewalls, as reported in\supercite{Shi2020Micromachines}.
The G1 grating had a duty cycle of $0.5$, and the grating lines were \SI{59}{\micro\metre} thick; The thickness of the remaining silicon substrate was \SI{240}{\micro\metre}.
A single tiled G1 grating was diced out from the wafer to a size of $\SI{203}{\milli\metre} \times \SI{75}{\milli\metre}$.

\paragraph*{Specimen}
The female breast specimen was a human breast tissue from an adult autopsy after a body donation for research (ethical agreement KEK-2012\_554).
It was without any grossly visible pathology, and was fixed in \SI{10}{\percent} buffered formaldehyd.
The specimen was vacuum-sealed in a plastic bag and placed in a cylindrical PMMA container with \SI{100}{\milli\metre} outer and \SI{90}{\milli\metre} inner diameter.
The container was filled with water to avoid air gaps.

\paragraph*{Measurement protocol}
During the CT scan, the specimen revolved continuously at \SI{1}{\rpm}
for five full rotations while the frames were acquired at \SI{20}{\hertz}.
After each rotation, the G0 grating was shifted by one sixth of its period.
The five-rotations protocol was repeated ten times with the \SI{10}{\milli\ampere} tube current, and three with \SI{2.5}{\milli\ampere}, and was interleaved with a reference phase-stepping measurements with the sample out of the beam.
The ten-repetition scan took \SI{1.5}{\hour} of wall-time.

\paragraph*{Dose estimation}
The term \emph{dose} was used to indicate the mean absorbed dose to the breast, approximated as the absorbed dose to a
0.25--0.75 volumetric mixture of the ICRU44 glandular and adipose tissues\supercite{ICRU44}, homogeneously distributed in a PMMA cylinder with a \SI{100}{\milli\metre} outer diameter and a \SI{90}{\milli\metre} inner diameter.
The absorbed dose was estimated by the means of Monte Carlo simulations (\texttt{GEANT4}\supercite{Agostinelli2003}), where the simulation geometry and source parameters were validated through measurements using BeO optically stimulated luminescence dosimeters (OSLDs) and LiF:Mg,Ti thermoluminescence dosimeters (TLDs)\supercite{Yukihara2022}.
The OSLDs and TLDs were calibrated in dose-to-water using a ISO N-60 photon field (mean energy \SI{47.9}{\keV})\supercite{ISO4037-1:2019} to approximate the mean energy of the X-ray field used for imaging (mean energy \SI{43.6}{\keV}).
The photons in the simulation were sampled from a spectrum approximating the X-ray tube in the experiment.
The source was collimated to match the extent of the absorption gratings G0 and G1 outlined in Fig.\,\ref{fig:principle}.
In a first experiment, the OSLDs and TLDs were placed upstream of the PMMA cylinder on the beam axis to establish a conversion factor between the simulated dose-per-primary to the absorbed dose to the dosimeters for a \SI{10}{\minute} irradiation.
In a second experiment, which served to validate the Monte Carlo model, the TLDs were placed on both external sides of the cylinder, upstream and downstream.
The measured doses to water (\SI{145}{\milli\gray} upstream and \SI{11.1}{\milli\gray} downstream for a \SI{10}{\minute} irradiation at \SI{10}{\milli\ampere}) were in agreement with the simulated dose-to-water in volumes matching the luminescence detectors (\SI{151}{\milli\gray} upstream and \SI{10.8}{\milli\gray} downstream), thus validating the Monte Carlo model parameters.
The Monte Carlo model was then used to score the dose to the homogeneous 0.25--0.75 volumetric mixture of the ICRU44 glandular and adipose tissues placed in the PMMA cylinder.
The volumetric fractions were established with a threshold-based segmentation of the reconstructed volumes.
The mean dose to the tissue mixture placed in the PMMA cylinder was calculated to be $\SI{21.9 \pm 2.4}{\milli\gray}$, which corresponds to a \SI{5}{\minute}-long CT measurement series at \SI{10}{\milli\ampere}.

\paragraph*{Data processing}
The sinograms corresponding to the five rotations with different G0 positions $x^j$ were stacked and we performed a signal-retrieval with a linear least-squares fitting of a sine to find the phase $\phi_i$, visibility $v_i$ and intensity $I_i$ in each $i$th pixel:
\begin{equation}
    I_i^j = \frac{I_i}{2}\, \left( v_i \sin\left( \frac{2 \pi}{p}\,x^j - \phi_i \right) + 1 \right)\ .
\end{equation}
We used an overarching least-squares optimisation to find the best-fit period of the sine $p$ common to all pixels.
The reference measurements, acquired between the tomography scans, were analysed in the same way to obtain the reference maps $\phi^r_i$, $v^r_i$ and $I^r_i$.
We then constructed the attenuation $p^I_i$ and differential-phase-contrast (DPC) $p^\phi_i$ sinograms taking as the reference the average of the two adjacent reference scans:
\begin{equation}
    p^I_i = -\log \left( \frac{I_i}{I^r_i} \right), \quad p^\phi_i = \phi_i - \phi^r_i \ .
\end{equation}
The attenuation sinogram $p^I$ was corrected for beam-hardening effects, for which we used a separate measurement of PMMA slabs of different thicknesses.
We further applied to the attenuation sinogram a ring-removal algorithm based on a combined wavelet-FFT filter with damping of 1, 3 wavelet transform levels and a db5 wavelet filter\supercite{Munch2009stripe}.
The attenuation volume was reconstructed with the FDK algorithm, for the phase-contrast we first used the Hilbert filter and then back-projection.
In both cases we used the ASTRA Toolbox GPU implementations\supercite{VanAarle2015}, cone-beam geometry and the voxel size of \SI{85.68}{\micro\metre}.
The reconstructed attenuation volume was treated with a TomoPy implementation of a reconstruction-space ring-removal algorithm ($\theta_\text{min} = 80$, $\text{threshold} = 0$)\supercite{Gursoy2014tomopy}.
Calibration to HU and HUp units\supercite{Donath2010TowardsClinical} was done by setting air to $-1000$ and a water region to zero.
The fused images were obtained by first normalising the phase-contrast reconstructed volume such, that the grey-levels of the adipose and glandular tissues, measured in three manually-selected ROIs each, corresponded to the ones in the attenuation volume.
Then, the attenuation volume was low-pass-filtered and the phase one high-pass-filtered with an in-plane Gaussian kernel of $\sigma = \SI{1.5}{\pixel}$. The resulting volumes were added.

\paragraph*{Quantitative analysis}
The reconstructed volume slices were obtained by averaging a varying number of sinograms before the reconstruction.
For the images corresponding to the doses \SIrange[range-phrase=--]{5.5}{16.5}{\milli\gray}, the average of 1--3 series with the current of \SI{2.5}{\milli\ampere} was used.
For the \SIrange[range-phrase=--]{22}{222}{\milli\gray} ones, 1--10 series with \SI{10}{\milli\ampere} were used.
Three circular regions containing the adipose and three with the glandular tissue were selected.
They are marked in Fig.\,\ref{fig:CT_images}.
The CNR was estimated by the average contrast between the tissue types and the standard deviation in the adipose regions.
For each image, we found numerically the minimal FWHM size of an isotropic 3D Gaussian kernel necessary to reach a CNR of 5\supercite{rose2013vision}.
The value was chosen based on the Rose criterion, which states that a CNR of 5 is sufficient to detect features.
We considered only the point spread function (PSF) introduced by the filtering, which is system-independent and sets the lower limit on the resolution. The total resolution of the imaging system is also influenced by the PSFs of the source and the detector.

\paragraph*{Derivation of the ideal-gratings limit}
We estimated the limit of the performance of the system assuming ideal gratings in a numerical Fresnel wave-propagation simulation.
The focal spot size of the X-ray source and its spectrum were considered by propagating accordingly weighted source fields.
We have modelled gratings with ideal, defect-free lamellae, but otherwise their geometry and material content, to the best of our knowledge, corresponded to reality.
The visibility in the model was \SI{17.6}{\percent}, which we interpret as the upper limit for the performance of the interferometer with this geometry.
The sensitivity of a GI-CT system to refraction increases linearly with the visibility and, further, the dose requirement inversely with the square of the sensitivity.
The increase of the visibility from \SI{9.4}{\percent} (currently achieved by our system) to \SI{17.6}{\percent} (theoretical limit) would then lower the dose requirement for the phase contrast by a factor of $3.5$.
We assumed that with the ideal gratings the attenuation would not change, so the intersection point of the attenuation and phase contrast best-fit lines would be at the resolution of \SI{417}{\micro\metre} and the dose of \SI{6.50}{\milli\gray}.
In Fig.\,\ref{fig:Relative_CNR}, we show the correspondingly shifted fused GI-CT curve.
To avoid extrapolation the limit does not extend beyond the measured points.



\printbibliography 

\section*{Acknowledgements}
The authors are grateful to Gordan Mikuljan and Philipp Zuppiger of PSI for their fantastic technical expertise and support.
The authors acknowledge the clean room facilities of PSI and the technical staff for the support in gratings fabrication.
This work has been funded by the SNF R’Equip grant 206021\_189662 (SiDRY), the ETH-Research Commission Grant Nr.\ ETH-12 20-2, an ETH Doc.Mobility Fellowship, the Promedica Stiftung Chur, the SNF Sinergia Grant Nr.\ CRSII5 183568, the PHRT-Pioneer Project Nr.\ 2021-612 CLARINET as well as the Swisslos Lottery Fund of Kanton Aargau.

\section*{Author contributions}
Z.W., M.R., M.\,Stampanoni and M.\,Stauber conceptualised the GI-CT system;
M.R. designed and built the system and conceptualised the measurement;
M.R. and A.P. analysed the data with contributions from S.S., S.v.G., J.X., M.P. and M.-C.Z.;
S.S. and M.R. set up the control system;
S.S. and A.P. implemented the wave-propagation simulation;
L.R., K.J. and Z.S. manufactured the G1 grating;
E.Y. and J.C. estimated the dose;
Z.V. provided the breast specimen;
R.A.K.-H., C.L., T.N., Z.V. and A.B. provided clinical expertise;
M.R. wrote the manuscript with contributions of all authors.

\section*{Competing interests}
M.\,Stauber is the CEO and a co-founder of GratXray~AG, Z.\,Wang is a co-founder of GratXray~AG, M.\,Stampanoni is a member of the BoD and a co-founder of GratXray~AG, A.\,Boss is a member of the BoD of GratXray~AG, L.\,Romano is the acting CSO of GratXray~AG and M.\,Rawlik is the acting CTO of GratXray~AG.

\clearpage
\onecolumn

\section*{Supplementary material}

\begin{figure*}[h!]
  \centering
  \includegraphics{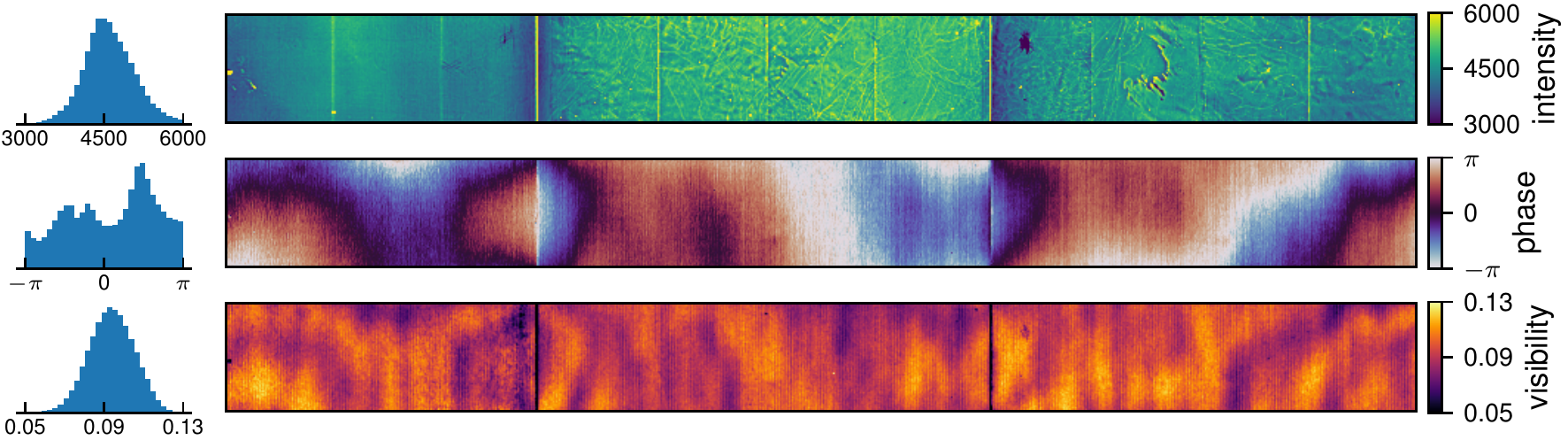}
  \caption{
    \textbf{The reference phase-stepping scan without the sample in the beam.}
    \emph{Top}: The intensity profile and its histogram.
    \emph{Middle}: The phase profile.
    \emph{Bottom}: The visibility profile. The average visibility is 0.094, the best regions are 0.12.
  }\label{fig:reference-profile}
\end{figure*}

\begin{figure}[h!]
  \centering
  \includegraphics[width=0.48\linewidth]{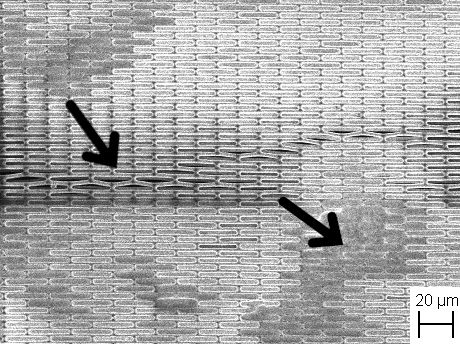}
  \caption{
    \textbf{SEM image of the surface of the G0 grating.}
    The gold lamellae were electroplated in high-aspect-ratio cavities in polymer (brighter on the image). A slight overplating defect is indicated with the arrow on the right-hand side. The arrow on the left side indicates where the lamellae detached from the polymer forming a gap. The grating was manufactured by Microworks GmbH, Germany.
  }\label{fig:gratings}
\end{figure}

\end{document}